\documentclass[aps,prb,twocolumn,superscriptaddress]{revtex4}
\usepackage{amssymb}
\usepackage{graphicx}
\usepackage{amsmath}
\usepackage{amsfonts}
\usepackage{amssymb}%

\begin{document}
\title{Anharmonic phonons in few layer MoS$_2$: Raman spectroscopy of ultra low energy compression and shear modes. }

\author{Mohamed Boukhicha}
\affiliation{Universit\'{e} Pierre et Marie Curie, IMPMC, CNRS UMR7590, 4 Place Jussieu, 75005 Paris, France}
\author{Matteo Calandra}
\email[]{matteo.calandra@upmc.fr}
\affiliation{Universit\'{e} Pierre et Marie Curie, IMPMC, CNRS UMR7590, 4 Place Jussieu, 75005 Paris, France}
\author{Marie-Aude Measson}
\affiliation{Laboratoire Mat\'{e}riaux et Ph\'{e}nom\`{e}nes Quantiques UMR 7162 CNRS, Universit\'{e} Paris Diderot-Paris 7, 75205 Paris cedex 13, France}
\author{Ophelie Lancry}
\affiliation{HORIBA Jobin Yvon S.A.S., 231 rue de Lille, 59650 Villeneuve d'Ascq}
\author{Abhay Shukla}
\email[]{abhay.shukla@upmc.fr}
\affiliation{Universit\'{e} Pierre et Marie Curie, IMPMC, CNRS UMR7590, 4 Place Jussieu, 75005 Paris, France}

\begin{abstract}
Molybdenum disulfide (MoS$_2$) is a promising material for 
making two-dimensional crystals and flexible electronic and optoelectronic devices
at the nanoscale\cite{Wang2012, Radisavljevic2011, Castellanos2012,Kim2012}.
MoS$_2$ flakes can show high mobilities and have even been integrated in nanocircuits 
\cite{Radisavljevic2012, Wang2012b}. A fundamental requirement for
such use is efficient thermal transport. Electronic transport generates 
heat which needs to be evacuated, more crucially so in nanostructures. 
Anharmonic phonon-phonon scattering is the dominant intrinsic
limitation to thermal transport in insulators. 
Here, using appropriate samples, ultra-low energy Raman spectroscopy and first principles calculations,
we provide a full experimental and theoretical description of compression and shear
modes of few-layer (FL) MoS$_2$. We demonstrate that the
compression modes are strongly anharmonic with a marked
enhancement of phonon-phonon scattering as the number of layers
is reduced, most likely a general feature of nanolayered materials
with weak interlayer coupling.
\end{abstract}

\maketitle

\section{Introduction}

Bulk MoS$_2$ is made of vertically stacked layers (single formula unit consisting of a Mo sheet sandwiched between two S sheets) weakly
held together by Van der Waals forces, with $2$ layers  per unit cell. While MoS$_2$ is an indirect gap semiconductor, it displays a crossover
to a direct gap semiconductor with a resulting marked  increase of
photoluminescence upon reduction of crystal thickness down to one
layer\cite{Mak2010}. 

In the case of a layered hexagonal system, the shearing modes are twofold degenerate as there are two 
equivalent in-plane shear-directions. Generally an N-layer flake has
$2*(N-1)$ shearing modes and $N-1$ compression-extension (noted compression henceforth) modes. Thus, while single layer MoS$_2$ 
should have neither one nor the other mode, in FL MoS$_2$ additional modes should appear with respect to bulk or the bilayer which both have 2 degenerate shear
modes and one compression mode.
\begin{figure}
\includegraphics[scale=0.5,angle=0]{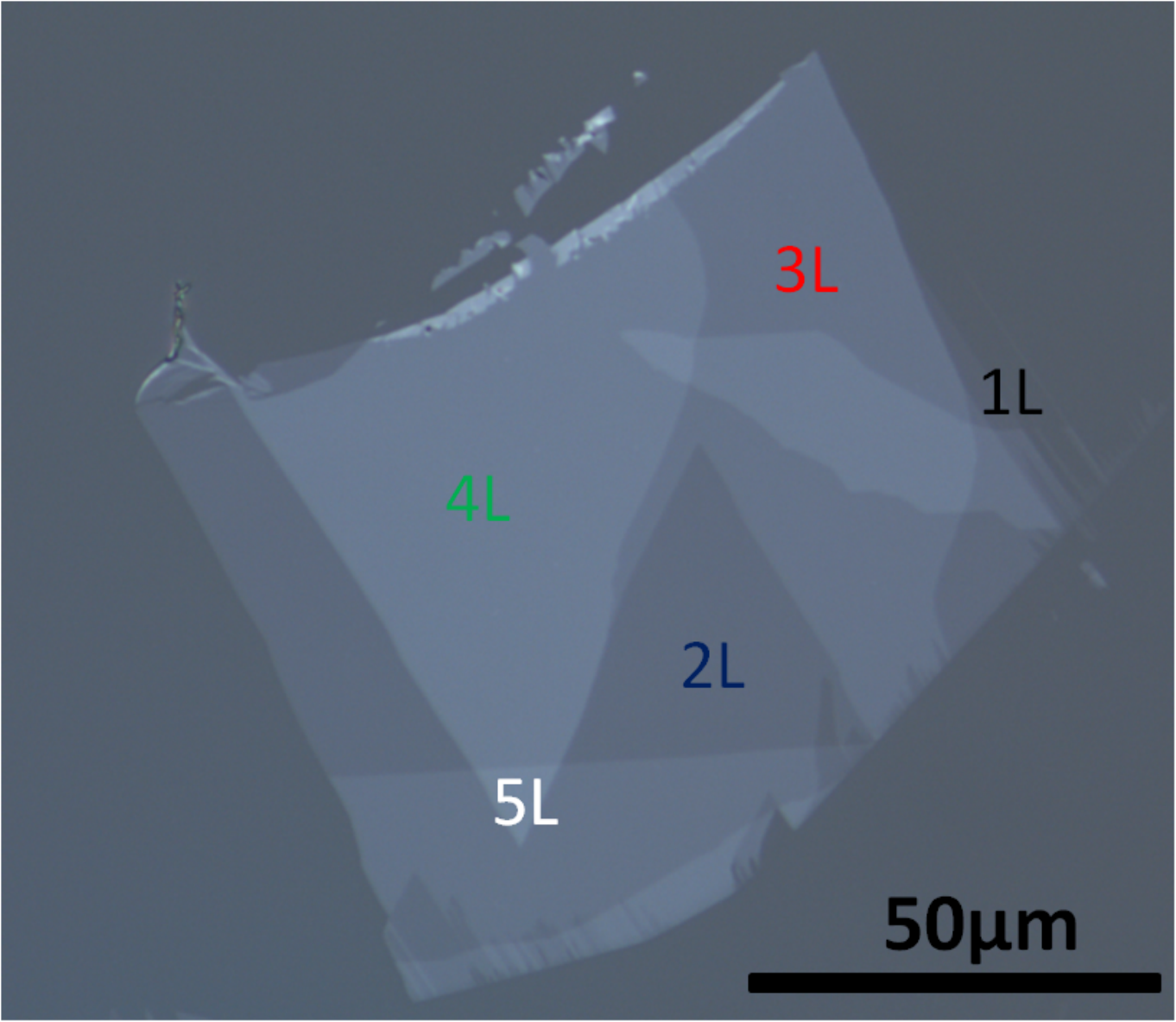}
\caption{The few-layers MoS$_2$ flake made by
 anodic bonding and used for the experiment. 
The number of layers is indicated.}
\label{flake}
\end{figure}

Both modes, importantly, are expected at very low energies since the interlayer interaction in MoS$_2$ is
weak ($\omega_{\nu}< 55$ cm$^{-1}$ or $74$K). Therein lies their relevance to transport, both thermal and electronic.
At room temperature these phonon modes are all thermally populated and influence thermal transport
via phonon scattering with defects and impurities or via phonon-phonon
scattering. As anharmonicity is an intrinsic mechanism, in clean
samples it is the dominant limitation to the thermal conductivity. 
Determining the behaviour of these modes as a function of the flake thickness is
of the greatest importance for nanoelectronic devices based on MoS$_2$.

Recently low energy Raman modes in Graphene \cite{Marzari2012} and
MoS$_2$\cite{Plechinger2012,Zeng2012,Zhang2013} were measured for FL flakes on p-doped Si
substrates. In Ref. \onlinecite{Plechinger2012} only a single shear mode was detected and
no compression modes were seen. 
Low energy modes are weak in intensity
and, to eliminate a broad low-energy background due to inelastic scattering of free carriers in
the p-doped substrate\cite{Chandrasekhar1980}, these spectra were
recorded in crossed geometry (i.e. the polarization of
the outcoming light is perpendicular to that of the incident beam) 
where all compression modes are forbidden.
In Ref. \onlinecite{Zeng2012},  beside the shearing mode detected in 
Ref. \onlinecite{Plechinger2012}, a second feature was detected. 
No polarization analysis of the Raman spectra was performed and one
of the features was attributed to a compression mode 
from a fit to a $1/N$ behaviour ($N=$ number of layers). 
In Ref. \onlinecite{Zhang2013} a detailed study of shearing and
compression modes up to 19 layers has been carried out. 
The authors were able to classify the low energy Raman
peaks in two groups, namely those that stiffen with increasing $N$ 
and those that soften with increasing $N$. Then using a chain model
they were able to build fan diagrams and obtain MoS$_2$  shearing and
compression strength. No first principles calculations of the Raman
spectra were carried out.  

\begin{figure*}[t]
\centerline{\includegraphics[scale=0.42,angle=0]{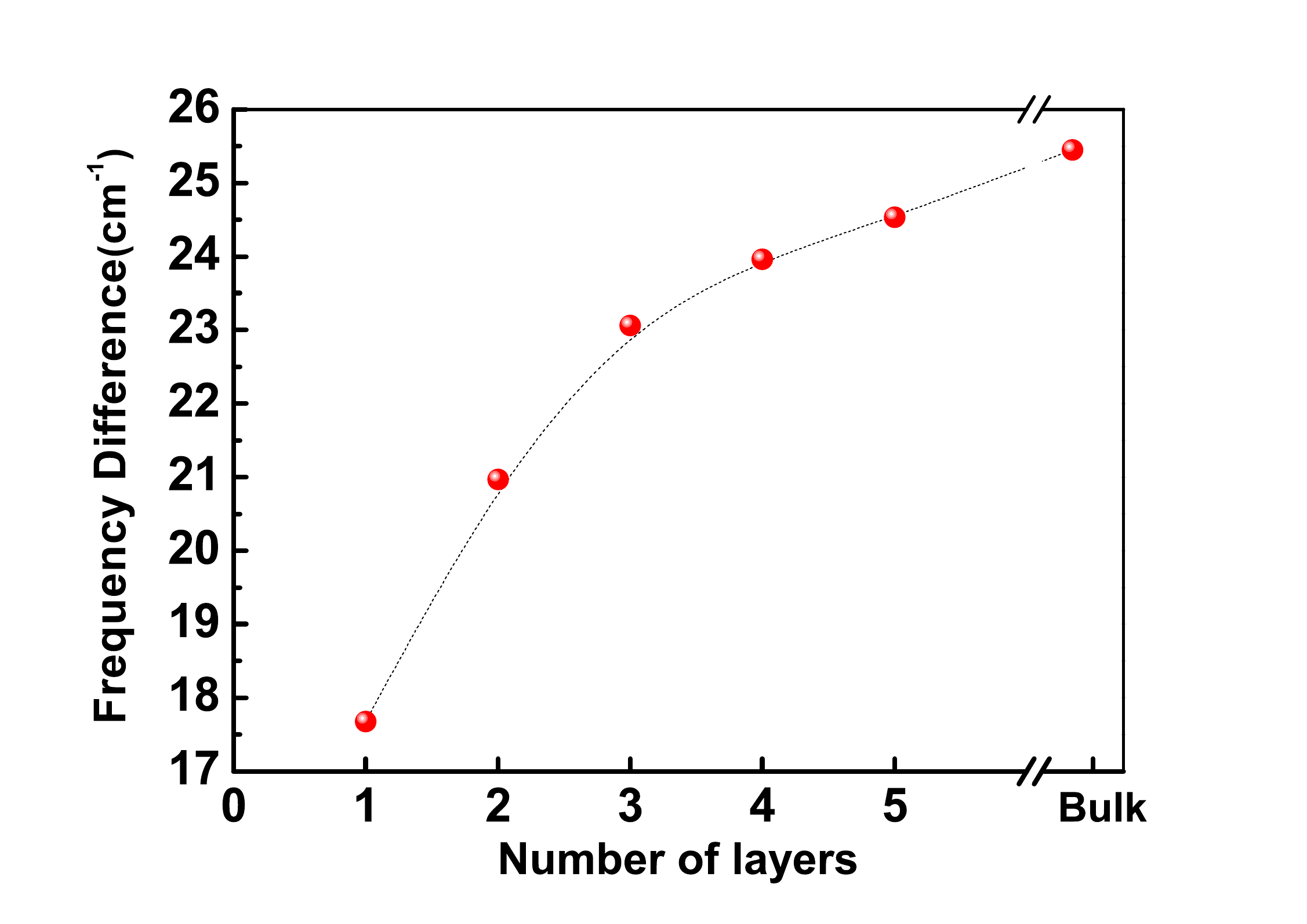}\includegraphics[scale=0.32,angle=0]{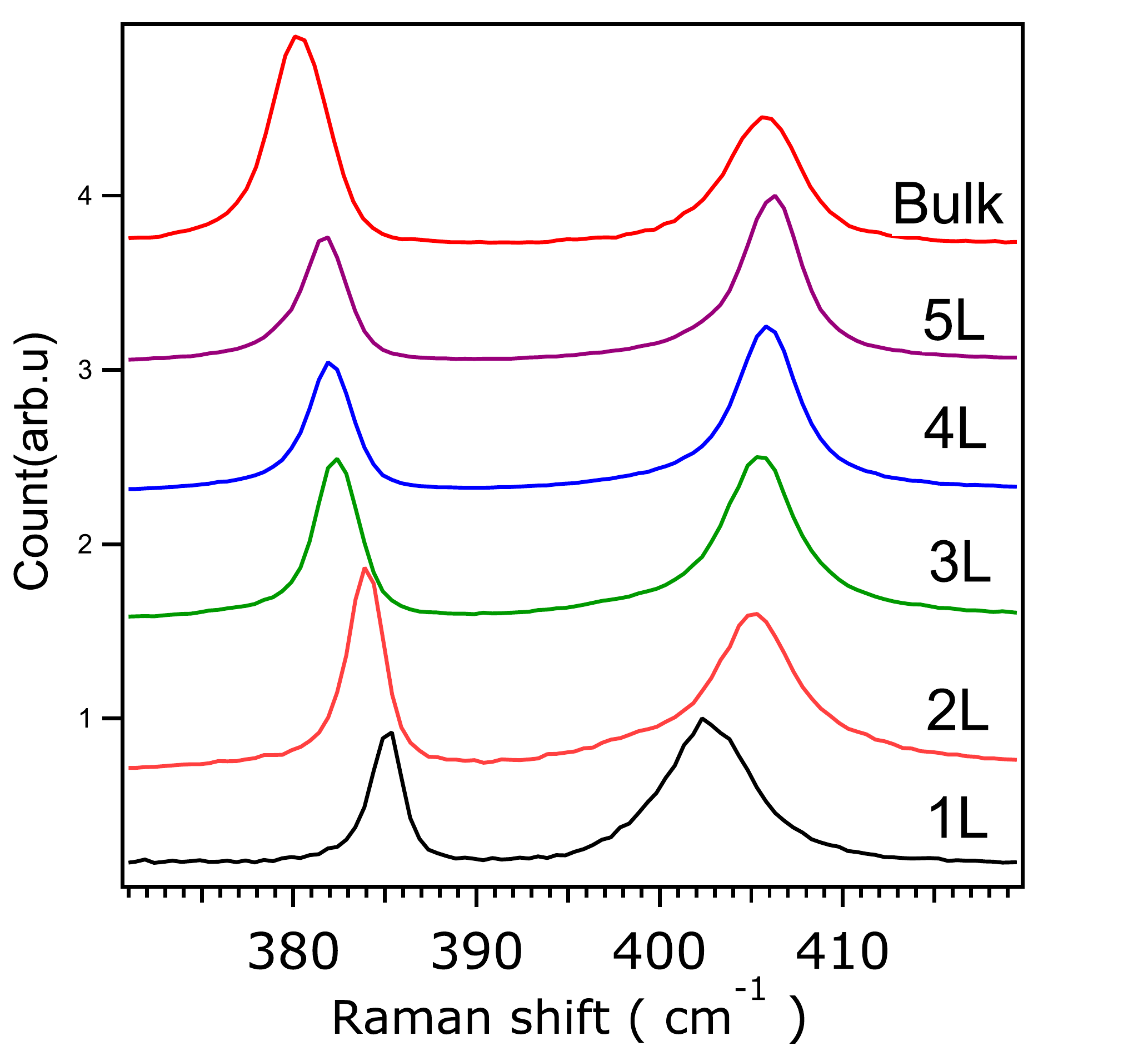}}
\caption{Energy difference of the E$_{2g}$ and the A$_{1g}$ phonon
  modes at high energy versus the number of layers (left) and
  experimental high-energy Raman spectra as a function of the layer
  number (right).}
\label{fig:Rthickness}
\end{figure*}

In this work we measure low-energy Raman spectra as a function of
the number of layers in multilayer MoS$_2$.   We overcome the 
difficulty related to the presence of a broad low-energy background 
due to inelastic scattering of free carriers in
the p-doped substrate\cite{Chandrasekhar1980} by using FL MoS$_2$
on borosilicate glass substrates.
We measure
shear and compression modes from 1 to 5 layers. By performing first
principles calculations of the position\cite{QE} and
intensity\cite{Lazzeri} of Raman peaks we obtain a complete 
understanding of shear and compression modes in FL MoS$_2$ . 
We also analyze theoretically the dependence of the main shearing mode as a function
of applied pressure and show that it behaves linearly at low pressure
(below 1GPa).

\begin{figure*}
\centerline{\includegraphics[scale=0.54,angle=0]{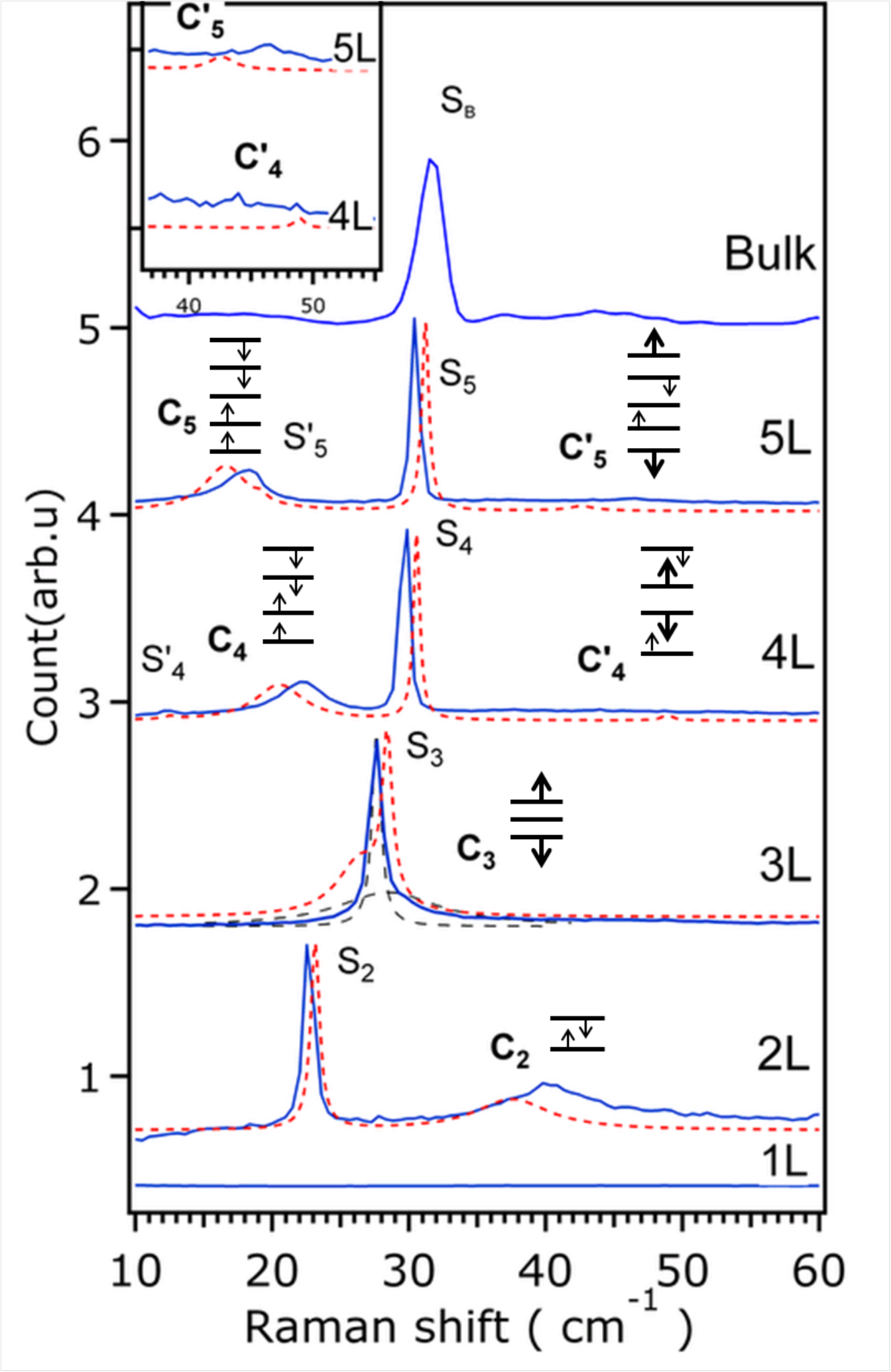}\includegraphics[scale=0.6425,angle=0]{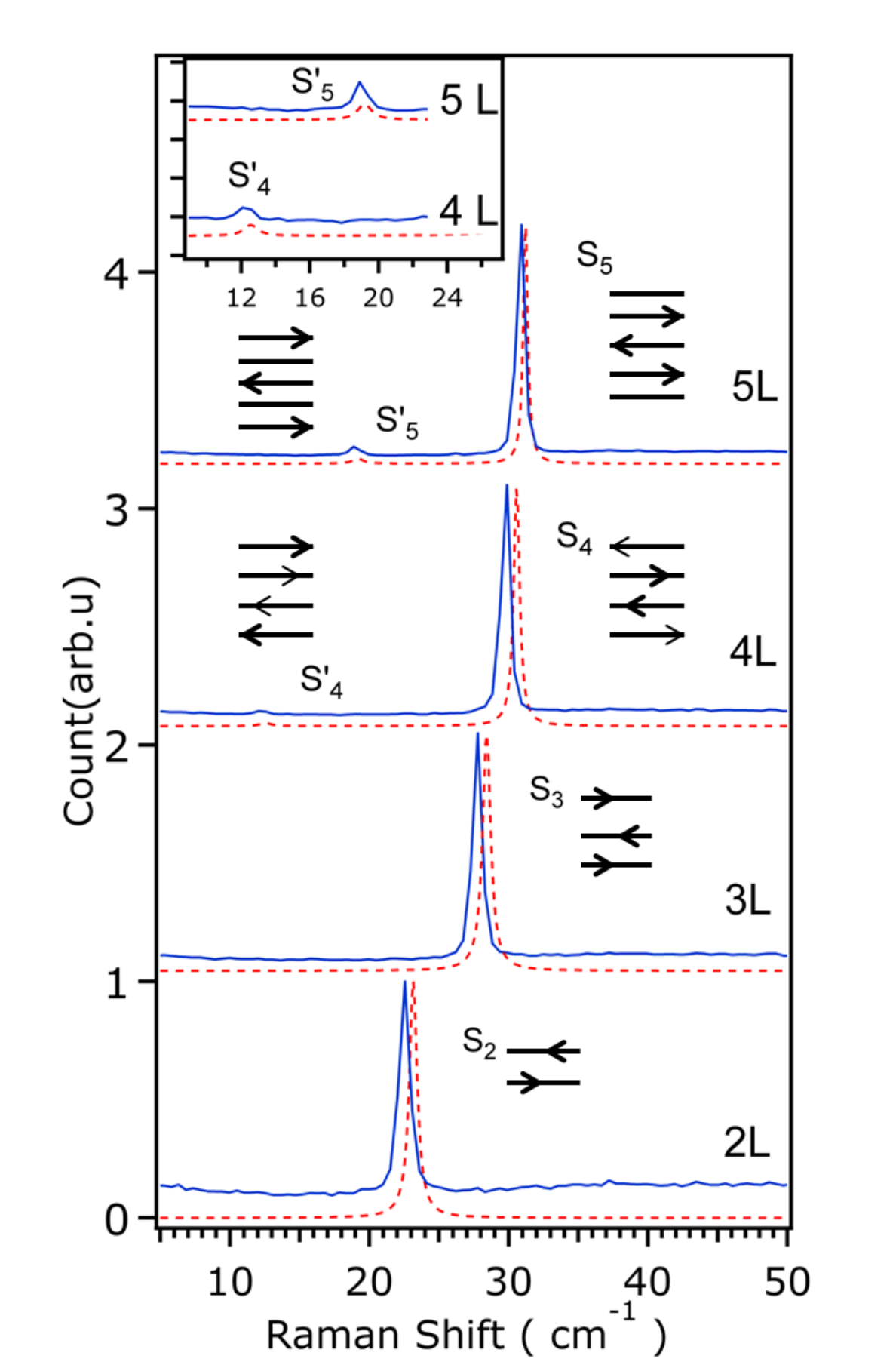}}
\caption{Experimental (blue) and
theoretical (red-dashed) Raman spectra  in
parallel (left) and crossed (right)
configuration. The inset shows a blow up of the low energy
region. 
Also shown are the schematized MoS$_2$ layer movements for the compression (left) and the shear (right) modes. All arrows may be simultaneously reversed
and the size corresponds to the amplitude}
\label{fig:Raman}
\end{figure*}

\section{Experimental}

The samples are made using the
anodic bonding method which bonds a bulk flake to a borosilicate glass substrate \cite{shukla2009,gacem2012}
at a temeprtaure between 130°C and 200°C and a high voltage which may range from 200V to 1500V. The
flake is then mechanically cleaved leaving large few layer samples on
the glass (see Fig. \ref{flake}). The sample used in this work is exceptional in that it provides
all different thicknesses used on the same flake, making comparison easy.
The sample thickness was identified first by optical contrast
and then confirmed by atomic force microscope and Raman spectroscopy 
\cite{Lee2010}, as shown in Fig. \ref{fig:Rthickness}.

MicroRaman spectra  of
the FL MoS$_2$ are measured ($532$ nm radiation, $\approx 1$ mW laser power) in backscattering 
configuration with parallel and crossed polarization geometry. 
To enable measurements down to $\approx 10$ cm$^{-1}$ on a
single-grating spectrometer (LabRAM HR
from HORIBA Jobin Yvon), 
an ultra-low	wavenumber filtering (ULF$^{TM}$)
accessory for $532$ nm wavelength was used.
These volume Bragg gratings can be fabricated with diffraction 
efficiencies as high as 99.99\% and the linewidth
narrower than 1 cm$^{-1}$ at FWHM that corresponds to 3-4 cm$^{-1}$ cut-off 
frequency at -60 dB from maximum. They also make a
unique notch filter for Rayleigh light rejection by sequential 
cascading of several Bragg notch filters, enabling ultra-low
frequency Raman measurements with single stage spectrometers \cite{Glebov,Glebov2}
To avoid laser heating, a laser power of 1~mW was focused through a 50X or
100X microscope objectives.

In order to evaluate the actual temperature on the sample and to
exclude laser-heating, we measure for all thicknesses 
both Stokes and antiStokes features at high and low energy.
We then determine the temperature as
\begin{equation}
T=\frac{\hbar\omega}{k_B} \ln\left\{\frac{I_S}{I_{AS}}
 \left(\frac{\omega_L+\omega}{\omega_L-\omega}\right)^4\right\}
\label{eq:cardona}
\end{equation}
where $\omega$ is the Raman shift and $\omega_L$
the pulsation of the laser light. For a more precise determination of
the temperature we use the intensity of the high energy A$_{1g}$ and 
E$_{2g}$ modes. We obtain $T\approx 360 $K.
Thus the laser heating is $\approx 60$ K, i. e. negligible.

\section{Theory}

Calculations were performed by using density functional theory in the local density approximation\cite{pz}.
The  QUANTUM-ESPRESSO\cite{QE} package was used with norm-conserving
pseudopotentials and a plane-wave cutoff energy of $90$ Ry. Semicore
states were included in Mo pseudopotential.
All calculations were performed at 0 and 6 Kbar uniaxial pressure,
corresponding to the uniaxial pressure imposed in the anodic bonding procedure. The
crystal structure at a given pressure is obtained by keeping the same
in-plane experimental lattice parameter as in bulk samples. The interlayer distance
is obtained by imposing a 6 Kbar pressure with respect to the bulk
experimental structure at 0 Kbar. The distance between the plane is then kept constant
for all N-layers flakes, but Sulfur height, the only free parameter,
is optimized following internal forces.
Phonon frequencies, born-effective charges and Raman tensor were
calculated using a $8\times8\times1$ k-point grid for the monolayer
and a $8\times8\times8$ for the bulk. Using this grid, phonon frequencies of shearing
and compressing modes are converged with an accuracy of  $1$
cm$^{-1}$.
Raman intensities were
calculated with the method of Ref. \onlinecite{Lazzeri} in the Placzek
approximation. 

The intensity of a mode $\nu$ is written as
$$I^{\nu}\propto I_{0} ^{\nu}\, (n_\nu+1) /\omega_\nu$$
 where $\omega_\nu$ and 
$n_\nu$ are the phonon frequency and the occupation of the phonon mode
$\nu$. Moreover, $I_0^{\nu}=|{\bf e}_i\cdot {\bf A} {\bf e}_o|^2$ where
${\bf A}$ is the Raman tensor while $ {\bf e}_i$ 
and ${\bf e}_o$ are the the polarization of the incident and scattered
radiation respectively. In table \ref{tab:intensity} we give the calculated
value of $I_0^{\nu}$ as a function of the layer number and the 
experimental geometries for the low energy modes.
\begin{table}\label{tab:intensity}
\begin{tabular}{c c c}
N=2   & & \\ 
$\omega_\nu ($cm$^{-1})$ &${\bf e}_i \parallel {\bf e}_o $ &   ${\bf e}_i \perp {\bf e}_o $  \\\hline
     23.1                        &   0.282           &  1.0     \\
     37.6                        &   1.0                &  $<$0.001   \\ \hline
N=3   & & \\ 
$\omega_\nu ($cm$^{-1})$ & ${\bf e}_i \parallel {\bf e}_o $  &  ${\bf e}_i \perp {\bf e}_o $  \\\hline
     16.31          &    $<$0.001    &   $<$0.001  \\  
     26.43          & 1.0    &   $<$0.001   \\ 
     28.42          &  0.73 &   1.0 \\
     45.95          &   $<$0.001     & $<$0.001      \\ \hline
N=4   & & \\ 
$\omega_\nu ($cm$^{-1})$ &  ${\bf e}_i \parallel {\bf e}_o $  &  ${\bf e}_i \perp {\bf e}_o $  \\\hline
 12.56    &     $<$ .005   &    $<$0.001 \\
 20.53    &     0.69      &     $<$ 0.001    \\
  23.34   &        $<$0.001        &            \\
  30.57   &        1.0        &   1.0      \\
  47.45   &        $<$0.001        &     $<$0.001       \\
  48.90   &       $<$0.001        &     0.13   \\  \hline
N=5   & & \\ 
$\omega_\nu ($cm$^{-1})$ &  ${\bf e}_i \parallel {\bf e}_o $  &
${\bf e}_i \perp {\bf e}_o $  \\\hline
10.04   &   $<$0.001    &  $<$0.001 0  \\
16.61   &   0.42 &   $<$0.001 \\  
19.18   &   0.02  &  $<$ 0.002  \\
26.50   &  $<$0.001      &  $<$0.001   \\
31.21   &  1.0    &  1.0   \\
31.25   &  $<$ 0.001 &  $<$0.001  \\
42.62    &  0.17   &  $<0.001$  \\
50.45   &    $<0.001$       &  $<0.001$  \\
\end{tabular}
\caption{Raman intensity $I_0^{\nu}$ for different modes $\nu$ and as
  a function of the number
of layers $N$ for backscattering geometry (labeled $\parallel$ geom.)
and cross cackscattering geometry. The intensity is normalized to the
most intense low energy mode (below $100$ cm$^{-1}$ ). Note
that in the two and three layer case, the compression mode has
stronger intensity $I_0^{\nu}$ then the main shear mode, however the
larger linewidth suppresses its $I^{\nu}$. }
\end{table}

The experimental spectrum is then obtained as
\begin{equation}
I(\omega)\propto \sum_\nu I^{\nu} \delta(\omega-\omega_\nu)
\end{equation}
In order to compare with experiments, the Dirac $\delta$ functions
are convoluted with the experimental linewidths.

At ambient pressure, the calculated frequencies are in excellent
agreement with previous calculations \cite{Wirtz}, however they disagree
with the calculations of  Ref. \onlinecite{Zeng2012}. In particular,
for a MoS$_2$ bilayer at zero uniaxial pressure and using the 
experimental in-plane lattice parameter 
we find $20$ cm$^{-1}$
for the main shearing mode. Performing structural optimization of both in-plane and
interlayer distance (keeping always the same empty region between
periodic images) we find $24.49$ cm$^{-1}$ for the shear mode.
Thus the shear phonon
frequency weakly depends on the choice of the experimental or
theoretical in-plane lattice parameters in the calculation.
This has to be compared with $22$ cm$^{-1}$ in 
our experimental Raman data, with $\approx 23$ cm$^{-1}$ in experimental Raman data of
Ref. \onlinecite{Zeng2012} and with  $\approx 19.5$ cm$^{-1}$ in
Raman data of Ref. \onlinecite{Plechinger2012}.
In Ref. \onlinecite{Zeng2012} the shearing mode was calculated at
$35.3$ cm$^{-1}$ using the theoretical lattice structure and the LDA
approximation which would correspond to a very large applied pressure, not relevant to the experiments in consideration.

\section{General discussion on secondary shear and compression modes}

As already stressed in the introduction, a MoS$_2$ N-layer flake has
$2*(N-1)$ twofold degenerate shearing modes and $N-1$ compression
modes. 
Shear modes correspond to rigid layer
displacements perpendicular to the c-axis.
The twofold degeneracy depends on the crystal symmetry of the lattice.
In MoS$_2$ for example it is equivalent to rigidly shift a subset of
layers with respect to one in-plane crystalline axis or the other.
For an N layer flake (with N$<5$) the rigid layer displacement patterns are
schematically illustrated in Fig. \ref{fig:Raman} (right).
The hardest shear mode always corresponds to the
rigid shift of the innermost layers, these being more
tightly bound by the bilateral interaction with the
other layers. The hardest (also named principal or primary) 
shear mode is labeled S$_n$, where $n$ is the number of layers.
The shear mode $S_n$ is Raman active and is also the most
intense of all the shear modes. 
As the number of layer increases, there are
more possibilities of rigidly shifting layers. For example in a 
three layer flake, it is possible to shift only the top (or bottom)
layer keeping the other two fixed. This mode is however softer
then the main shear mode, as the outer layers have only one nearest 
neighbour layer with weaker binding.
In our case and at 6 Kbar uniaxial pressure, the secondary shear
mode is calculated to be at 16.3 cm$^-1$, but with essentially 
zero Raman intensity (although this mode is not forbidden
by symmetry). In the general case of an N-layer flake, there are
$(N-1)$ independent ways of shifting a subset of layers with respect
to the others and their energies lie between that of the (softest) secondary 
shear mode related to the shift of extremal layer and the (hardest) main
shear mode. 
 
Compression (or extension) modes are rigid vibrations of the layers in the direction
perpendicular to the layers. For an N layer flake 
(with N$<5$) the rigid layer displacement pattern are
schematically illustrated in Fig. \ref{fig:Raman} (left). 
As in the case of shear modes it is possible to identify a primary or
main compression mode. The main compression mode is the one
corresponding to (i) the lower half of the layers shifting in the same
direction and (ii) the higher half of the layers in the opposite direction.
The main compression mode (labeled C$_n$)  is the softest compression
mode. Secondary compression modes arise when the top half  (respectively
bottom half) layers are not all displaced in the same
direction. Secondary compression modes are higher in energy then the
main compression mode (see Fig. \ref{fan}, the rigid layer model of \onlinecite{Michel2012} and the discussion below).
 
\section{Results}

In Fig. \ref{fig:Raman} we show measured and calculated spectra in
parallel and crossed geometries. The peaks are
normalized to the main shear mode intensity and the theoretical
spectra are convoluted by the experimental linewidth.

By comparing the main shear-mode (labeled S$_n$) to existing measurements
\cite{Plechinger2012},
we notice that in our samples this mode is systematically harder by
$\approx 3$ cm$^{-1}$. We attribute this to the anodic bonding method
which binds flakes to a glass substrate electrostatically due to
the creation of a space charge in the substrate. This also generates
an electrostatic pressure on the flake bound to the substrate. Knowing the depth of
this space charge layer (1-2 microns) this uniaxial pressure along the c-axis \cite{Anthony1983}
can be estimated to be in the range of 3-6 Kbar. 

We thus performed first principles calculations as a function of
pressure and obtain an essentially linear behaviour of the
main shearing mode for a MoS$_2$ bilayer as a function of uniaxial pressure, with a 
linear coefficient that is $1.17$ cm$^{-1}$/Kbar (the second order coefficient
in the fit is 0.04 cm$^{-1}$/Kbar$^2$). 
Comparing the measured spectra with the calculated ones for $6$ Kbar uniaxial pressure, we find remarkable agreement. 

\begin{figure}
\includegraphics[scale=0.5,angle=0]{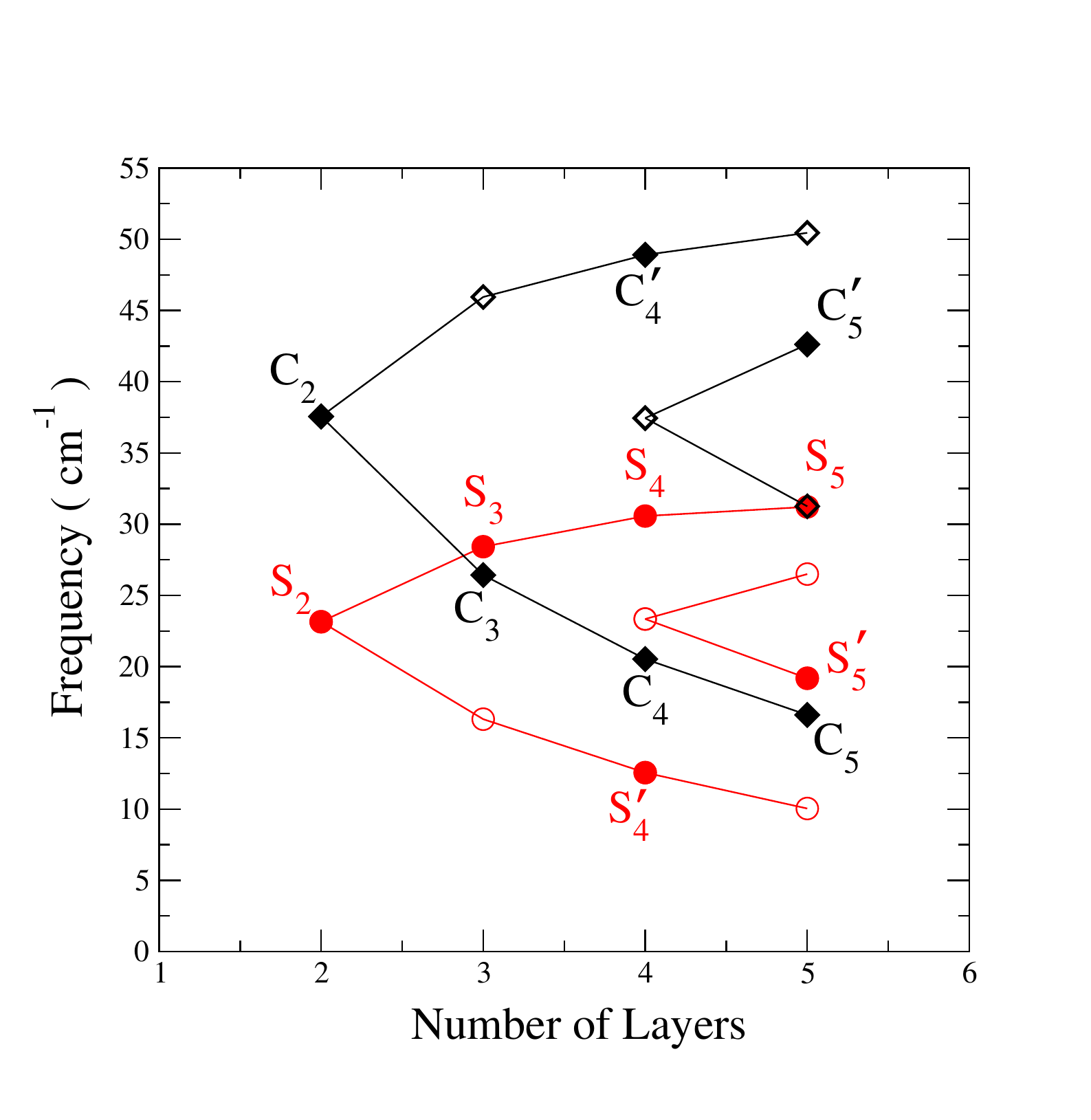}
\caption{
Fan diagram of calculated shear (red) and compression (black)
mode frequencies as a function of the number of layers.
Full symbols represent modes visibles in our Raman experiments
in parallel and crossed configuration. The labels of the different
modes are the same as in 
Fig. \ref{fig:Raman}. }
\label{fan}
\end{figure}

Besides the main shear mode, already detected in
Refs. \onlinecite{Plechinger2012, Zeng2012} we measure secondary shear
(S$_n$ and S$_n^{\prime}$)
and compression (labeled C$_n$ and C$_n^{\prime}$) modes 
(see Figs. \ref{fan} and \ref{fig:Raman} ).  
In Ref. \onlinecite{Zeng2012} only the $C_n$ compression 
mode was detected. 

The compression mode is clearly visible in 2, 4 and 5 layer samples
and less so in the 
3 layer sample because it coincides in energy with the shear mode. 
In the 4 and 5 layer samples, theory also accounts for the
additional shear (S$_n^{\prime}$) and compression (C$_n^{\prime}$) modes detected in experiments.

\begin{figure}
\includegraphics[scale=0.6,angle=0]{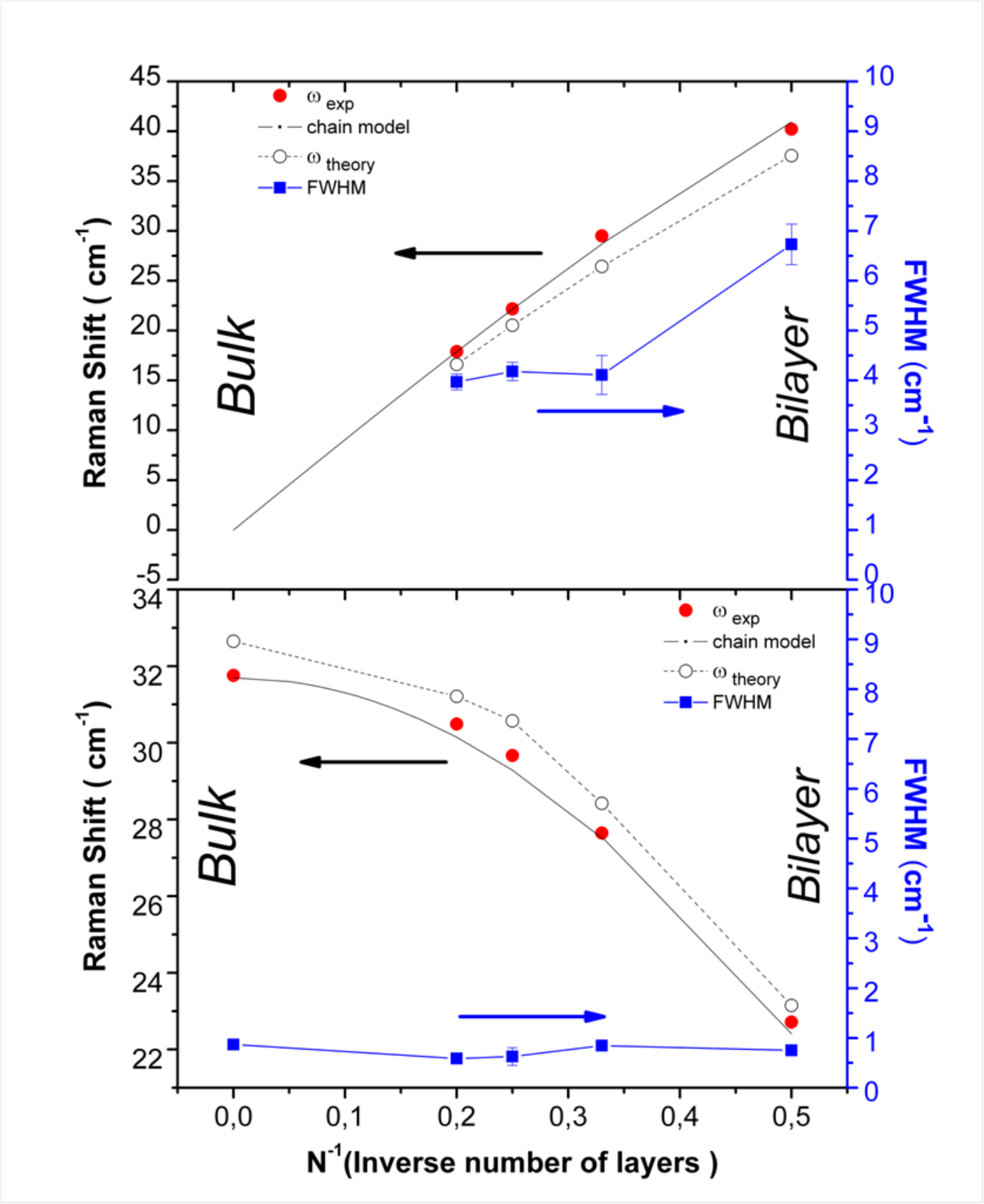}
\caption{Energy and linewidth ($\gamma$) full-width half-maximum
of compression (top) and shear (bottom) modes as a 
function of the inverse layer number.}
\label{energywidth}
\end{figure}

It is worthwhile to recall that
the energy of shear and compression modes can either increase or
decrease as a function of the number of layers $N$, as shown in the fan
diagrams in Ref.  \onlinecite{Michel2012} and in Fig. \ref{fan}.
In FL MoS2, the energy of Raman visible E$_g$ shear modes
increases with the number of layer, while that of Raman visible
$A_{1g}$ compression modes decreases.

The qualitative behaviour of the position of the main shear and compression modes
as a function of layer number can be easily understood in a simple rigid-layer (or chain) model\cite{Ashcroft}. 
We write for the $S_n$ shear-mode frequency
\begin{eqnarray}
\omega_{S}= \frac{1}{\sqrt{2}\pi c} \sqrt{\frac{\alpha_S}{\mu}}\sqrt{1+\cos\left(\frac{\pi}{N}\right)}
\label{eq:freqshear}
\end{eqnarray}
where $\mu=30.75 \,{\rm Kg}/{\rm m}^2$ is the rigid layer mass per
unit and $\alpha_S$ is the shearing strength. The C$_n$ compression mode
behaves as
\begin{eqnarray}
\omega_{C}=  \frac{1}{\sqrt{2}\pi c} \sqrt{\frac{\alpha_C}{\mu}}\sqrt{1-\cos\left(\frac{\pi}{N}\right)}
\label{eq:freqcompr}
\end{eqnarray}
and $\alpha_C$ is the compression strength.
From the experimental curves we obtain $\alpha_S=27.44 \times 10^{18}
{\rm N}/{\rm m}^3$, twice the value in graphene \cite{Marzari2012, Michel2012}, and 
$\alpha_C=420.44 \times 10^{18}
{\rm N}/{\rm m}^3$.
Both these effects are in part explained by the smaller interlayer distance in MoS$_2$
(the Sulfur-Sulfur distance along c is $3.03 \AA$) with respect to graphite
($3.35 \AA$). 
The agreement between Eqs. \ref{eq:freqshear}, \ref{eq:freqcompr} and
experimental data is shown in Fig. \ref{energywidth}, validating the 
chain model and the extracted values of $\alpha_S$ and $\alpha_C$.

In Fig. \ref{energywidth} we also plot the variation of the linewidth
$\gamma$ for the S$_n$ shear and  the C$_n$ compression 
modes as a function of the number of layers $n$. Shear
mode linewidth is resolution limited while all the compression modes are very broad
(roughly 7 times broader), the
linewidth of the C$_2$ mode being the largest.
As the broadening is inversely proportional to the
phonon-phonon scattering time, our result indicates that the phonon 
scattering time of compression modes is approximately $7$ times
smaller than that of shear modes. Thus the contribution of
optical modes to the intrinsic thermal conductivity of MoS$_2$ flakes 
is dominated by scattering to compression modes.
In a MoS$_2$ bilayer, the scattering time of compression modes is
nearly 9 times smaller then that of shear modes.

The compression modes linewidths are generally larger then those
of shearing modes as the potential is more anharmonic for a
displacement perpendicular to the MoS$_2$ layers then for a shearing displacement. 
In the case of C$_2$, the linewidth is enhanced  with respect to C$_n$ with $n>2$ 
due to the fact that more  channels for anharmonic decay are available. 
Indeed for $n>2$, the compression mode is lower in energy 
(or at the same energy for $n=3$) then the shearing mode. 
As such it can only decay into two  acoustic modes of opposite momentum. 
In the case $n=2$, the compression mode is at roughly twice the energy 
of the shearing mode S$_2$. Thus the compression mode can decay into 
(i) two acoustic modes of opposite momentum,
 (ii) an acoustic and a shearing mode of opposite momentum,
or (iii) two shearing modes of opposite momentum.

\section{Conclusion}

In conclusion we have measured primary and secondary shearing and
compression modes in MoS$_2$ from 1 to 5 layers. 
The compression modes are found to be strongly anharmonic, 
with phonon-phonon scattering increasing upon reducing the number 
of layers. Thus compression modes represent the overriding 
optical phonon contribution to the intrinsic thermal
conductivity of MoS$_2$ flakes, a crucial 
aspect of any use of these in future nano or microelectronic devices.
The relevance of our work is far reaching as compression modes
are most likely strongly anharmonic in all flakes obtained
from weakly-interacting layered-materials such as
few layer graphene, transition metal dichalcogenides and 
topological insulators. In all these systems a crucial limit
to thermal transport could be the anharmonicity of compression phonon modes.

\section{Acknowledgements}

We acknowledge K. Gacem for help in sample preparation. 
This work was supported by French state funds managed by the 
ANR within the Investissements d'Avenir programme under reference  
ANR-11-IDEX-0004-02 and ANR-11-BS04-0019.
Calculations were performed at the IDRIS supercomputing center.

%Unused bibitems

%

%

\end{document}